\documentclass[
	a4paper, 
	10pt, 
	unnumberedsections, 
	twoside, 
]{LTJournalArticle}

\usepackage{todonotes}

\usepackage[nolist]{acronym}

\usepackage{amssymb}

\runninghead{} 

\footertext{\textit{RISC-V Summit Europe, Munich, 24-28th June 2024}} 

\title{Enhanced LPDDR4X PHY in 12\,nm FinFET} 

\author{%
	Johannes Feldmann\thanks{Corresponding author: \href{mailto:j.feldmann@rptu.de}{\tt j.feldmann@rptu.de}}\,, Jan Lappas, Mohammadreza Esmaeilpour,\\ Hussien Abdo, Christian Weis, and Norbert Wehn
}

\date{\footnotesize{University of Kaiserslautern-Landau}}

\renewcommand{\maketitlehookd}{%
	\begin{abstract}
		\noindent
        The demand for memory technologies with high bandwidth, low power consumption, and enhanced reliability has led to the emergence of \acs{lpddr} \acs{dram} memory.
        However, power efficiency and reliability depend not only on the memory device but also on its interfacing.
        To enable advanced monitoring of \acs{lpddr} \acs{dram} devices and interface tuning, we propose a \acs{lpddr} \acs{phy} implemented in $12\,nm$ FinFET technology.
        A RISC-V subsystem offers software-controlled \acs{dram} interface access as well as external interfaces to connect additional sensors for monitoring temperature and current consumption of \acs{lpddr} \acs{dram} devices.
	\end{abstract}
}

\begin{document}

\begin{acronym}
\acro{lpddr}[LPDDR4X]{Low Power Double Data Rate 4X}
\acro{phy}[PHY]{Physical Layer}
\acro{dfi}[DFI]{DDR PHY Interface}
\acro{dma}[DMA]{Direct Memory Access}
\acro{spi}[SPI]{Serial Peripheral Interface}
\acro{uart}[UART]{Universal Asynchronous Receiver Transmitter}
\acro{soc}[SoC]{System on Chip}
\acro{dram}[DRAM]{Dynamic Random Access Memory}
\end{acronym}

\maketitle 

\vspace{-8pt}
\section{Introduction}
\vspace{-2pt}
Driven by emerging applications like machine learning there is a continuous demand for \ac{dram} with higher capacity, higher bandwidth, and lower power consumption.
\ac{lpddr} is one of the latest standards with a main focus on low power consumption, which makes it particularly suitable for mobile devices.


At the core of \ac{lpddr} memory interfacing, the \ac{phy} plays a crucial role in transferring data between the memory device and the memory controller.
To achieve a reliable connection to the \ac{lpddr} memory, the \ac{phy} needs to adjust all drivers and receivers by impedance calibration and delay trimming, also called read-write-leveling (training).
In state-of-the-art DDR~\acp{phy}~\autocite{10nm_lpddr4}, the calibration algorithms are realized in software running on efficient processor cores.
These cores are either proprietary~\autocite{Synopsys} or open-source, like the RISC-V IBEX core~\autocite{Ibex} as part of the Wavious LPDDR PHY~\autocite{Wavious}. 
The software needs direct control over all \ac{lpddr} command, address, and data signals to perform the device initialization and read/write leveling (training sequence).

\begin{figure}[!b]
    \centering
    \vspace{-6pt}
    \includegraphics[width=0.80\columnwidth]{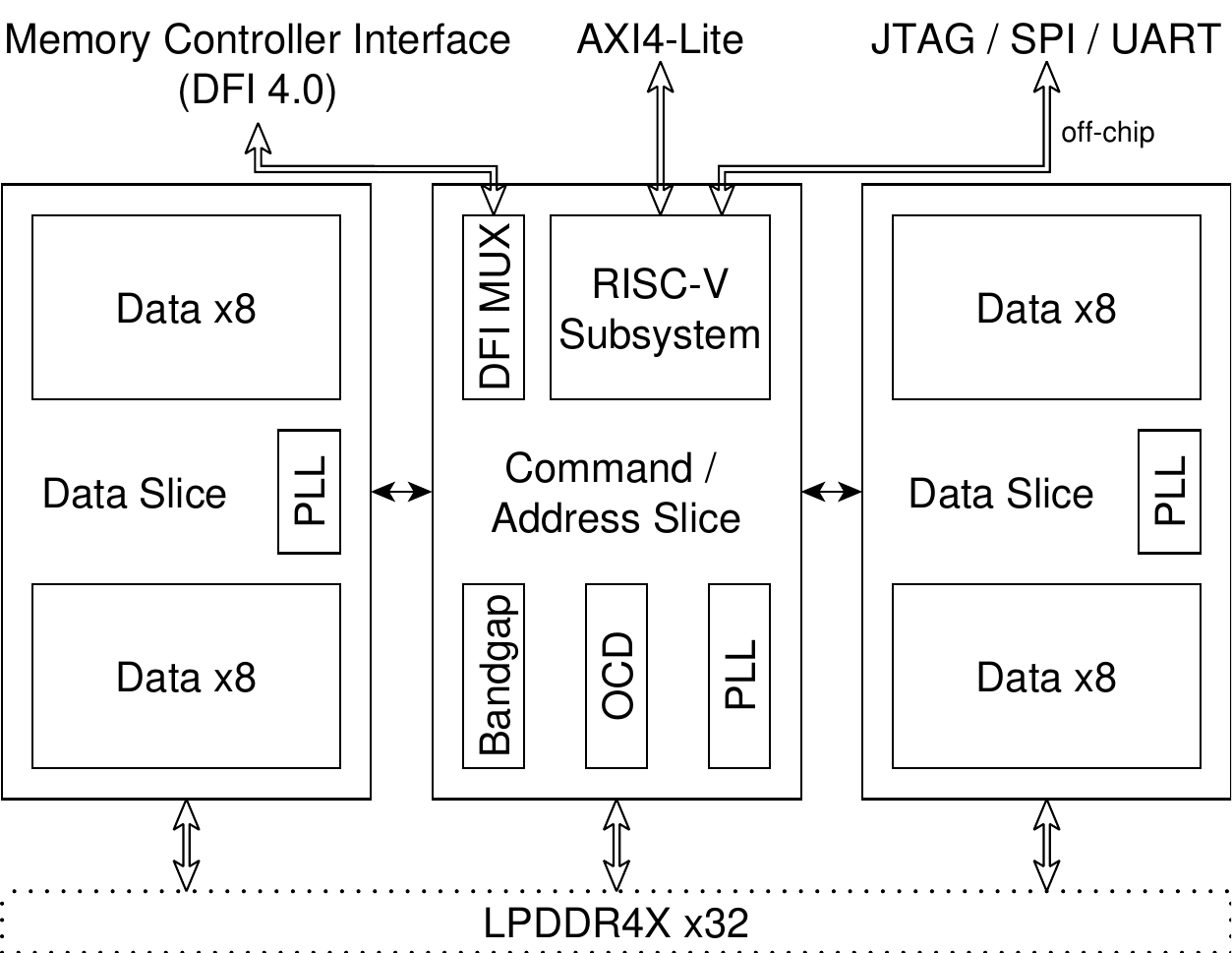}
    \caption{LPDDR4X PHY Architecture}
    \vspace{-12pt}
    \label{fig:phy_overview}
\end{figure}

However, state-of-the-art \acp{phy} feature \ac{dram} interface accessibility focus only on read/write leveling.
Additional tasks such as advanced monitoring or interface tuning are not supported.

In this paper, we enhance the state-of-the-art \ac{phy} architecture with software-controlled low-level \ac{dram} access using a \ac{dfi} Bridge and interfaces for off-chip sensors that are controlled by a RISC-V Subsystem.


\vspace{-8pt}
\section{Architecture}
\vspace{-2pt}

The \ac{lpddr} \ac{phy} is implemented in \textit{12\,nm} FinFET technology and has a maximum frequency of \textit{2133\,MHz}.
It consists of three modules, two Data Slices and one Command/Address Slice which form a 32-bit wide (x32) PHY used to connect two \ac{lpddr} channels, as shown in Figure~\ref{fig:phy_overview}.


\begin{figure}[!b]
    \centering
    \vspace{-6pt}
    \includegraphics[width=0.90\columnwidth]{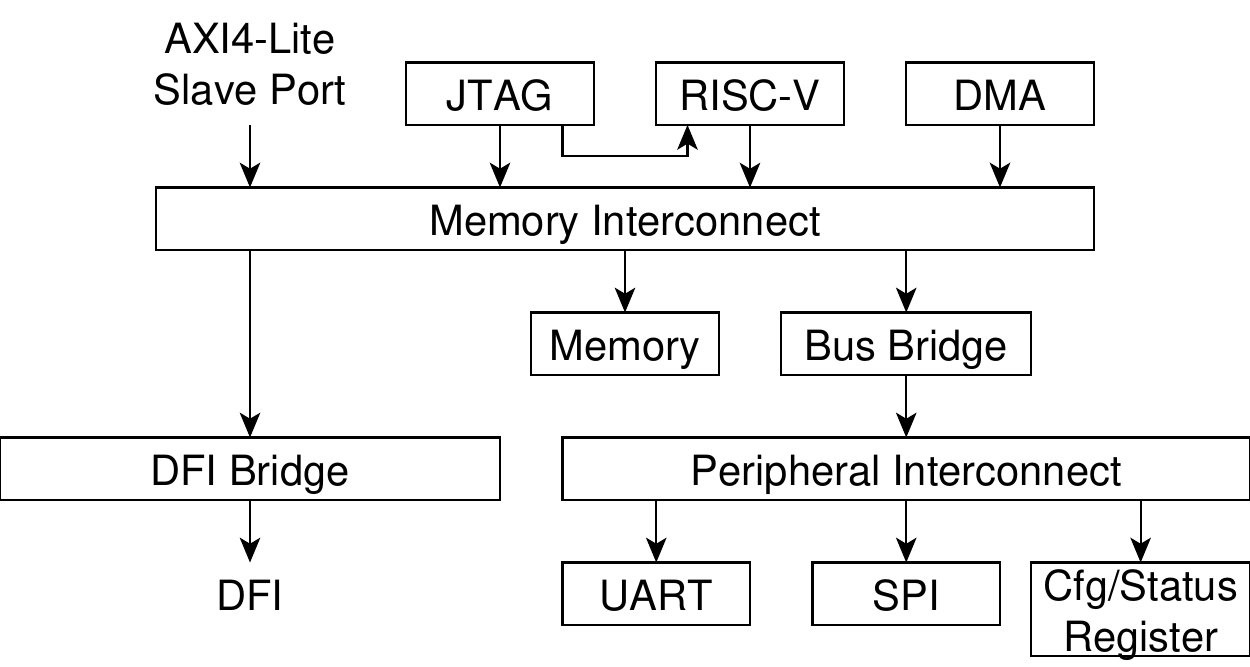}
    \caption{RISC-V Subsystem Architecture}
    \vspace{-12pt}
    \label{fig:phy_subsystem}
\end{figure}

\begin{figure*}[!t]
    \centering
    \vspace{-6pt}
    \includegraphics[width=0.75\textwidth]{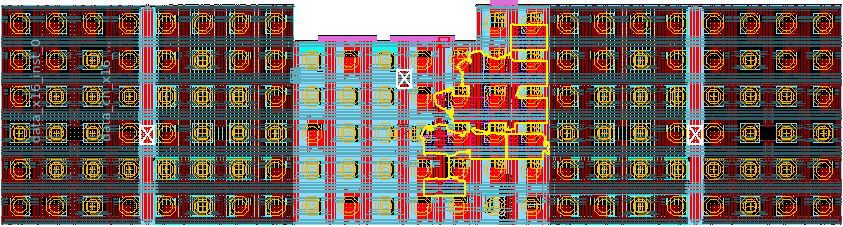}
    \caption{\ac{lpddr} \ac{phy} Layout. Yellow outline marks the RISC-V Subsystem including the DFI Bridge.}
    \vspace{-12pt}
    \label{fig:phy_layout}
\end{figure*}
The architecture of the RISC-V Subsystem is shown in Figure~\ref{fig:phy_subsystem}.
The subsystem is clocked at \textit{1066\,MHz} with a ratio of 1:2 to the \ac{phy} clock to match the \ac{dram} controller frequency.
The area-efficient 32-bit RISC-V core implements the RV32IMC instruction set and is accessible via an IEEE 1149.1-2013 compliant JTAG TAP.
The 32-bit AXI4-Lite Memory Interconnect links the core, a \ac{dma} controller, and an AXI4-Lite slave port, to \textit{64\,kB} of SRAM memory, the \ac{dfi} Bridge, and the Peripheral Interconnect.
The Bus Bridge connects the AXI4-Lite bus to a lightweight peripheral bus, which does not support concurrent read and write accesses.
The \ac{uart} and \ac{spi} units implement the off-chip communication to e.g. sensors.
The Configuration and Status Registers are used to control and monitor all internal modules of the \ac{phy}.



\begin{figure}[!hb]
    \centering
    \vspace{-6pt}
    \includegraphics[width=0.92\columnwidth]{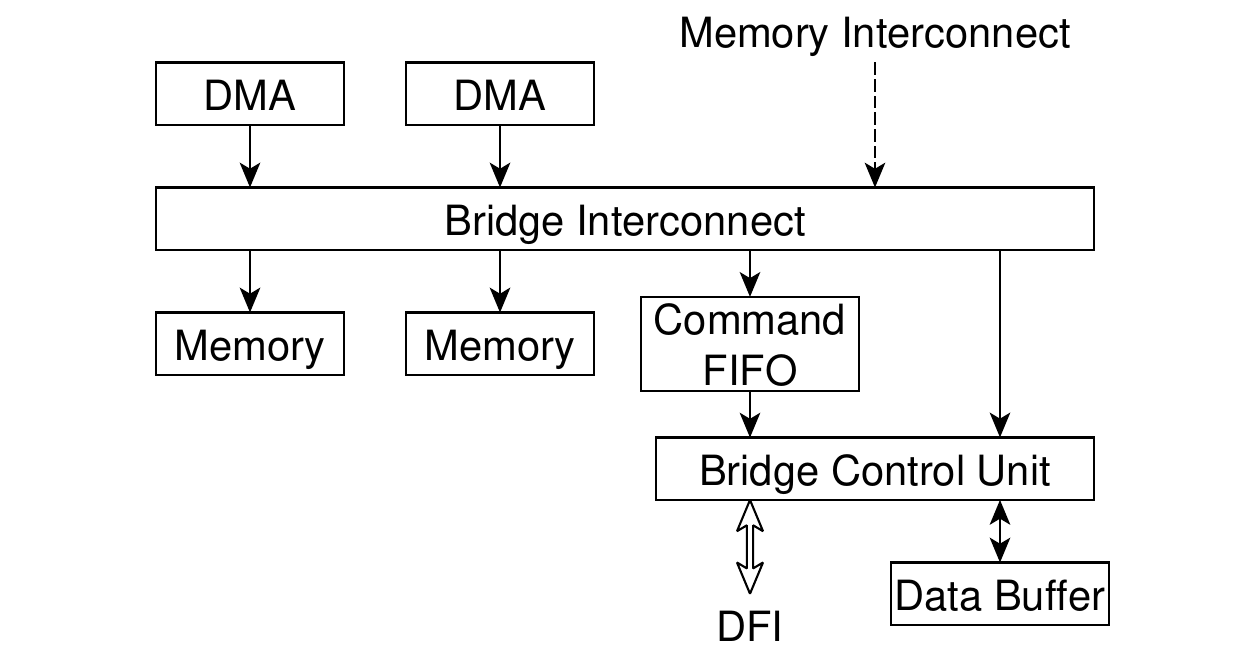}
    \caption{\ac{dfi} Bridge Architecture}
    \vspace{-12pt}
    \label{fig:phy_bridge}
\end{figure}

\begin{table}[!b]
    \centering
    \vspace{-6pt}
    \caption{Hierarchical Area}
    \label{tab:phy_area}
    \begin{tabular}{lrrr}
        \toprule
                                        &   \textbf{Cell [$\mu{}m^2$]} &  \textbf{SRAM [$\mu{}m^2$]}   \\
        \midrule           
        RISC-V Subsystem                & 24468                    & 106641                \\
        $\vdash$RISC-V                  & 4518                     & $-$                   \\
        $^{\llcorner}$DFI Bridge        & 4147                     & 48701                 \\
        \bottomrule
    \end{tabular}
    \vspace{-12pt}
\end{table}

The \ac{dfi} Bridge, shown in Figure~\ref{fig:phy_bridge}, consists of a 64-bit AXI4-Lite interconnect which connects two \ac{dma} controllers and the RISC-V core to two \textit{16\,kB} SRAMs, the Command FIFO, and the Bridge Control Unit.
The Bridge Control Unit manages the Data Buffer which contains 256 512-bit words and is used to store data of issued read and write commands.
To access the \ac{dfi} interface, a \ac{dfi} command needs to be issued to the Command FIFO.
Every \ac{dfi} command is a 64-bit word that contains two CA commands, timing information, and an index to a Data Buffer slot in case of a read or write command.
Both \ac{dma} controllers are used to move data between the memories and the command FIFO as well as the Data Buffer while the Bridge Control Unit is processing incoming commands.
Hence, the \ac{dfi} Bridge can issue continuously arbitrary command sequences while maintaining the maximum throughput of the \ac{dfi} interface.



\vspace{-8pt}
\section{Results}
\vspace{-2pt}

The \ac{lpddr} \ac{phy} presented is implemented in GlobalFoundries 12LPPlus technology using ARM Standard cells.
Synthesis is performed with the Synopsys DesignCompiler, and Place\&Route is carried out with the Synopsys IC-Compiler 2.
The SRAMs were generated using the ARM Artisan Memory Compiler.
The \ac{phy} has a size of \textit{2.89\,mm$^{2}$} and was taped out Nov. 2023. 
The layout is presented in Figure~\ref{fig:phy_layout} which shows the RISC-V subsystem with a yellow outline.
The corresponding area data is listed in Table~\ref{tab:phy_area}.

The total size of the \ac{phy} is determined by the number of bump pads.
The additional logic cells and SRAMs of the RISC-V subsystem can be hidden by the bump pads.
However, the additional interfaces for off-chip sensors need six additional bump pads that increase the area by \textit{0.135\,mm$^{2}$}.

The RISC-V Subsystem including all SRAMs consumes an average of \textit{12.6\,mW} of power at \textit{533\,MHz} and \textit{24\,mW} at the maximum frequency of \textit{1066\,MHz} under nominal conditions.
This is about \textit{15\,\%} of the overall power consumption of the \ac{phy}.

%

\vspace{-8pt}
\section{Conclusion}
\vspace{-2pt}

We presented an \ac{lpddr} \ac{phy} in \textit{12\,nm} FinFET technology that offers the RISC-V Subsystem software-controlled \ac{dfi} access and additional interfaces for external sensors.

In future publications, we will show the impact of this \ac{phy} on the overall system power consumption and reliability using the physical chip expected to be delivered in April 2024.

\vspace{-8pt}
\section{Acknowledgement}
\vspace{-2pt}

This work was funded by the European High-Performance Computing Joint Undertaking (JU) as part of the European PILOT project (No. 101034126).

\vspace{-8pt}

\printbibliography 

\end{document}